\documentstyle[aps,twocolumn,psfig]{revtex} 
\begin{document} 
\draft
 
\title{A Dynamic Approach to the Thermodynamics of Superdiffusion} 
\author{Marco Buiatti,$^{1}$ Paolo Grigolini$^{1, 2, 3}$ 
and Anna Montagnini$^{1}$} 
\address{$^{1}$Istituto di Biofisica del Consiglio Nazionale delle 
Ricerche, Via S. Lorenzo 26, 56127 Pisa, Italy} 
\address{$^{2}$Center for Nonlinear Science, University of North 
Texas, P.O. Box 305370, Denton, Texas 76203-5368} 
\address{$^{3}$Dipartimento di Fisica dell'Universit\'a di Pisa, Piazza 
Torricelli 2, 56100 Pisa, Italy} 
\maketitle 
\begin{abstract} 
We address the problem of relating thermodynamics to mechanics 
in the case of microscopic dynamics without a finite  
time scale.  
The solution is obtained 
by expressing the Tsallis entropic index $q$ as a function of the  
L\'evy 
index $\alpha$, and using dynamical rather than probabilistic  
arguments.  
\end{abstract} 
\pacs{05.20.-y, 05.45.+b} 
        The problem of establishing a connection between dynamics  
and  thermodynamics is still the object of wide interest among  
physicists. Recently rather than deriving 
Brownian diffusion from statistical mechanics, and consequently from  
thermodynamics, the inverse procedure was applied, of 
deriving Brownian motion from merely dynamical arguments  
\cite{[J95],[BMWG95]}. It was also shown that this is equivalent to  
deriving thermodynamics from mechanics. 
However, this mechanical derivation of thermodynamics rests on the assumption  
that a finite microscopic time scale exists. In the last few years an  
increasing interest has been devoted to generalized forms of Brownian 
motion that are incompatible with the existence of a finite  
microscopic time scale, notably 
the processes of superdiffusion, and especially the L\'evy diffusion 
\cite{[SZK93],[LEVY94],[KSZ96]}. Consequently, the dynamic derivation of these  
processes, at a first sight, seems to be incompatible with thermodynamics.  
 
It is well known \cite{[MS84]} that these  
diffusion processes, in the one-dimensional case, are characterized by  
probability distributions $p(x,t)$ whose Fourier transform in the  
symmetric case reads
\begin{equation} 
\label{pk} 
\hat{p}(k,t) = exp(-b|k|^{\alpha}t)  , 
\end{equation}       
where $\alpha$ is the L\'evy index ranging, in principle, in the interval  
$0 < \alpha < 2$, and $b$ denotes the diffusion intensity. Note that according to the analysis of  
\cite{[TFWG94],[AGW96]}, the processes corresponding to the more  
restricted range  
\begin{equation} 
\label{alfa1} 
1< \alpha < 2 
\end{equation} 
are compatible with dynamic derivations, including those of deterministic 
and Hamiltonian nature \cite{[ZSW93]}. The work of Ref. \cite{[AGW96]}  
illustrates a dynamic derivation resting on a stationary correlation  
function of the microscopic fluctuation, $\Psi_{W}(t)\propto  
1/{t}^{\beta}$ with $0 < \beta < 1$, namely with 
an infinite time scale, thereby making it impossible to derive  
thermodynamics as in Ref. \cite{[BMWG95]}. Here we show  
that in spite of this limitation, these dynamics are not incompatible with  
thermodynamics, provided that a non extensive rather than 
an extensive entropy is adopted.

Note that the diffusing system of Ref.\cite{[AGW96]}  
has a finite velocity with  
only two possible values, $W$  
and 
$- W$, corresponding to two distinct states of the system, and the time 
duration of these two states, $t$, is characterized by the probability 
distribution $\psi(t)$. The function $\psi(t)$, in turn, is  
proportional to the second time derivative of  $\Psi_{W}(t)$ 
and consequently $\psi(t) \propto 1/{t}^{1+\alpha}$ with $1<\alpha <2$. 
A jump of a given length $|x|$,  
with 
probability density $\Pi (x)$, is not instantaneously  
made,  
but it takes place through a time interval proportional to the distance  
$|x|$ travelled by the particle. Consequently, the correlation  
function with an infinite lifetime $\Psi_{W}(t)$ results in 
$\Pi(x) \propto  
1/{|x|}^{1+\alpha}$.

        A crucial aspect of this dynamic approach is that the distribution 
is characterized by finite moments, in spite of the fact that the moments 
of the L\'evy diffusion, from that of the second-order on, are divergent.  
The apparent conflict between these 
dynamically derived diffusion processes and the traditional L\'evy  
processes 
is settled by noticing that the dynamic derivation implies the existence 
of a propagation front, characterized by ballistic peaks  
\cite{[KZ93],[KZ94]}, 
and that the probability distribution of $x$  
between these peaks takes the shape of a  
truncated 
L\'evy process. The shape of an ideal L\'evy process is reached, instead,  
in the time asymptotic limit, as a consequence of the fact that the  
intensity of these peaks tends to zero, although very slowly. 
 
        It has to be pointed out that the historical foundation of the 
L\'evy 
processes is probabilistic in nature \cite{[note1]}. We note that,  
according to the perspective established by Jaynes \cite{[J57]}, there is  
a close connection between the second principle of thermodynamics and a  
probabilistic approach to statistical mechanics. This means that the  
thermodynamic foundation of the L\'evy diffusion should be more natural  
than its dynamic foundation. On the contrary, as  
recently pointed out by a group of authors \cite{[AZ94],[ZA95],[TLSM95]},  
the moment divergencies raise problems also to this thermodynamic  
foundation. These authors proved that if a 
proper use is made of the Tsallis entropy \cite{[T88]} 
\begin{equation} 
\label{tsentr} 
S_{q}[\Pi(x)] = - \frac{ 1- \int_{-\infty}^{\infty}{\Pi(x)}^{q}dx}{(1-q)} , 
\end{equation} 
the challenge posed by the infinite moments can be successfully 
addressed, since the constraint on the moments is set using ${\Pi(x)}^{q}$  
as a 
weighting function thereby ensuring convergence even where the standard 
weight would produce divergencies. Using this kind of weight, the L\'evy 
process can be derived by means of the standard procedure of entropy 
maximization resting on the technique of Lagrange multipliers. Later on in  
this paper 
we shall give more details on this approach. For the time being, let us  
limit ourselves to noticing that the maximization of the non-extensive  
entropy of Eq.  
(\ref{tsentr}) led the authors of 
Refs.\cite{[AZ94],[ZA95],[TLSM95]} 
to the conclusion that the entropic index $q$ is related to $\alpha$  
as follows: 
\begin{equation} 
\label{q1} 
q= \frac{3+\alpha}{1+\alpha} . 
\end{equation} 
 
        To realize the main purpose 
of establishing a connection between 
dynamics and thermodynamics, we must 
make the result of Eq. (\ref{q1}) compatible 
with  
the first steps 
successfully made by the authors of Refs. \cite{[LT98],[TPZ97],[CLPT97]} 
at the dynamic level.  
These authors laid the foundation for the connection between the entropic  
index $q$ and fractal dynamics. It is argued \cite{[TPZ97]} that it is  
convenient to 
generalize the definition of the Kolmogorov entropy, which is a sort of  
change per unit of time of the Gibbs entropy, by replacing the Gibbs  
entropy with the Tsallis entropy. The generalized form of Kolmogorov  
entropy, called $K_{q}$, is easily shown to yield the generalization of  
the well known Pesin identity \cite{[H94]} to 
\begin{equation} 
\label{kq} 
K_{q}={\lambda}_{q}, 
\end{equation} 
with ${\lambda}_{q}$ playing the role of a generalized Lyapunov  
coefficient. This important result is obtained by studying the distance  
$\Delta y(t)$  between two trajectories departing from two distinct  
initial conditions, at the distance $\Delta y(0)$ the one from the other,  
and assigning to $\delta (t)\equiv \lim_{|{{\Delta} y(0)}| \to  
0}\big{|}{\frac{\Delta y(t)}{\Delta y(0)}}\big{|}$ the following form: 
\begin{equation} 
\label{deltax} 
\delta (t)= (1+(1-q){\lambda}_{q}t)^{\frac{1}{1-q}} .  
\end{equation} 
It is evident that the traditional exponential sensitivity to the initial  
conditions is recovered from Eq. (\ref{deltax}) by setting $q=1$. Note  
that we adopt the symbol $y$ to denote \emph {microscopic}  
trajectories so as to avoid confusion with the \emph {probabilistic} and  
\emph {macroscopic}  
level of description, denoted by the symbol $x$. 
 
In Refs. \cite{[TFWG94],[AGW96]} it has been discussed how to derive 
superdiffusion on the basis of dynamic properties, 
with a procedure, valid also in the Hamiltonian case, which 
however recovers the results of the continuous random walk method  
of\cite{[KZ93],[KZ94],[ZK93],[ZK93D]} when applied to the same dynamic systems.  
These dynamic systems are bimodal maps, with two laminar  
regions, 
derived from that of Geisel and Thomae \cite{[GT84]}.  
With respect to the  
laboratory reference frame, the particles are characterized by the  
velocity $W$ or $- W$, according to whether they move in the right or left  
laminar region. The analytical form of the map describing the motion in  
the left laminar region is given by \cite{[GT84]}: 
\begin{equation} 
\label{map} 
y_{n+1}=y_{n} + a{y_{n}^{z}}, 
\end{equation} 
with $a>0$, $z>1$. This map extends from $y = 0$ to $y = d$  with the  
parameter $d$ defining the size of the laminar region. The right laminar  
region, with the same size $d$, is separated from the left by a chaotic  
layer and its analytical form is the same as that of Eq. (\ref{map})  
with $y$ denoting in this case the distance from the right border of  
the map. The particles starting from an initial condition  
very 
close to $y = 0$ are assumed to be driven by the continuous time 
approximation to Eq. (\ref{map}):  
\begin{equation} 
\label{referee} 
\frac{dy}{dt}=a{y}^{z}. 
\end{equation} 
Thus it is  
possible to evaluate the distribution of waiting times in  
the interval $(0,d)$. The assumption that trajectories  
are injected   
with uniform probability yields \cite{[GT84]} 
\begin{equation} 
\label{psi} 
\psi (t)=d^{z-1}a[1+d^{z-1}a(z-1)t]^{-\frac{z}{z-1}} . 
\end{equation} 
The right laminar region yields for the corresponding waiting time  
distribution the same analytical form as that of Eq. (\ref{psi}). Note that the  
research work of Refs. \cite{[TFWG94],[AGW96],[ZK93]} has established the  
connection between the index $z$ and the L\'evy index $\alpha$ to be given  
by 
\begin{equation} 
\label{alfa} 
\alpha=\frac{1}{z-1}. 
\end{equation} 
Note also that the value $z= 2$ is the border with the region where the  
first moment of $\psi(t)$ diverges, and represents the limit of validity  
of the dynamic treatment \cite{[AGW96]}. For values of $z < 3/2$, on the  
contrary, the second moment of $\psi(t)$, as well as the first, becomes  
convergent. Thus $z=3/2$ is the border between the L\'evy and the Gaussian  
region.  
 
        To assess the validity of the entropic prediction (\ref{deltax})  
we calculate $\delta(t)$         
using the dynamics of Eq. (\ref{map}) and we prove 
\begin{equation} 
\label{deltax2} 
\delta (t) = (1-a(z-1){y_{0}^{z-1}}t)^{-\frac{z}{z-1}} , 
\end{equation} 
where $y_{0}$ is the initial condition of one of the two trajectories under 
study. As done in \cite{[GT84]}, this initial condition is set so close  
to $y=0$ as to make it possible to adopt the 
continuous time representation of Eq. (\ref{referee}). Furthermore  
the assumption is made  
that $|\Delta y_n / y_n| << 1$. 
Notice that in the numerical calculations\cite{[KZ94]} it is usually  
set $d = 1/2$. Thus the  
divergence of $\delta(t)$ of Eq.(\ref{deltax2}) occurs at times much  
larger than the escape time from the  
laminar region. By comparing Eq. (\ref{deltax2}) with the general property  
of Eq. (\ref{deltax}), and using also Eq. (\ref{alfa}), we obtain 
\begin{equation} 
\label{q2} 
q=\frac{2+\alpha}{1+\alpha} 
\end{equation} 
and 
\begin{equation} 
\label{lambda} 
\lambda_q = a\frac{\alpha +1}{\alpha} y_{0}^{1\over\alpha} . 
\end{equation} 
This is an important result. However, to realize the main purpose 
of this paper, Eq. (\ref{q2}) must be made compatible 
with thermodynamics, in apparent conflict with 
the fact that it is seemingly at 
variance with Eq. (\ref{q1}), which results in fact from the 
adoption of thermodynamical arguments.  As a matter of fact, to  
establish a connection between the dynamic and the thermodynamic  
approach in the same spirit as that leading 
all the current attempts at unifying dynamics and  
thermodynamics, we must move from the canonical 
condition of \cite {[J95],[BMWG95]} to the  
microcanonical condition\cite{[note3]}. 
 
	To prove this interesting conclusion, let us 
give details on the variational approach of \cite{[ZA95]}.  
Let us consider the functional form $F_{q}(\Pi)$ defined by 
\begin{eqnarray} 
\label{fun} 
F_{q}(\Pi)\equiv  \frac{ 1- \int_{-\infty}^{\infty}{\Pi(x)}^{q}dx}{(1-q)}  
- \beta \int_{-\infty}^{\infty}|x|^{\nu} {\Pi(x)}^{q} dx \nonumber\\
+ \alpha  
\int_{-\infty}^{\infty} \Pi(x) dx . 
\end{eqnarray} 
This means that, according to the spirit of the technique of Lagrange  
multipliers, we plan to look for the entropy maximum under the constraint 
of keeping constant the $\nu-th$ moment, second term on the 
\emph{r.h.s} of Eq. (\ref{fun}), and the norm of the distribution  
$\Pi(x)$, 
third term on the  \emph{r.h.s} of Eq. (\ref{fun}). 
The authors of  
Refs.\cite{[AZ94],[ZA95],[TLSM95]} made the choice $\nu =2$.  
This choice was dictated by the fact that in their picture the  
diffusing particle undergoes a collision process at regular time  
intervals, of duration $\tau$, making the velocity variable $v$ fluctuate. Since 
$x = v\tau$, the constraint on the second moment of $x$ is actually a  
constraint on the first moment of the energy distribution. 
 
To establish a connection between entropic and dynamical arguments we  
are led to make a different choice. The space variable is made  
to change by the fluctuating  
velocity, which, as earlier pointed out, has only two 
possible values, $W$  and $-W$. During the motion in a laminar region, 
corresponding to the map of Eq. (\ref{map}), the velocity variable keeps  
the same value, either $W$ or $-W$. Consequently the space $x$ travelled  
is rigorously proportional to the time $t$ spent in a given laminar  
region, and the transition probability $\Pi(x)$ is related 
to the waiting time distribution $\psi(t)$ 
by the crucial relation 
\begin{equation} 
\label{pipsi} 
\Pi(x) = \psi(x/W)/W . 
\end{equation}  
This means that within the dynamic approach of this letter, the constraint 
has to be established on the first moment of $\Pi(x)$, this being  
essentially a 
constraint on the mean time spent in a given laminar region. In other  
words, this physical condition implies $\nu =1$. 
 
	By means of the entropy maximization, carried out by differentiating Eq.  
(\ref{fun}) with respect to $\Pi$, we get ($A$ is a function of  
$\alpha$ and both Lagrange parameters are determined by the  
corresponding moment values): 
\begin{equation} 
\label{pi2} 
\Pi(x)={A\over{[1+\beta(q-1)|x|^{\nu}]^{1\over{q-1}}}}. 
\end{equation} 
As argued by the authors of \cite{[TLSM95]}, we note that the L\'evy  
diffusion 
process is obtained for $N\to \infty$ from $\Pi(x,N)$ defined as the  
N-fold convolution product $\Pi(x)*\Pi(x)*...*\Pi(x)$  
\cite{[GK54]}. We notice that it is not necessary to assume the  
point of view that a transition controlled by $\Pi(x)$ takes place at any  
time step. In the dynamic case of the intermittent map of Eq. (\ref{map}),  
we can also follow a different approach. We define the mean waiting time  
in a laminar region, using as a weight the distribution $\psi(t)$. We call  
this time $T$ and we assume it to be finite. This is possible only if  
$\alpha > 1$, and this is the reason why our dynamic approach is confined  
to the interval $1 < \alpha < 2$. Then we consider a time $t >> T$. It is  
evident that for $t\to\infty$, the distribution $p(x,t)$ becomes very  
close to the N-fold convolution product $\Pi(x)*\Pi(x)*...*\Pi(x)$ with  
$N$ given by the integer approximating the quantity $t/T$. As done by the  
authors of \cite{[TLSM95]}, we apply the L\'evy-Gnedenko generalized  
central limit theorem \cite{[GK54]} and we get a L\'evy diffusion with  
the L\'evy index $\alpha$ determined by the condition that the transition  
probability $\Pi(x)$ fulfils the asymptotic limit 
\begin{equation} 
\label{lim} 
\lim_{t \to \infty} \Pi(x) \propto {1\over{|x|^{1+\alpha}}} . 
\end{equation} 
Comparing (\ref{lim}) to (\ref{pi2}), we are led to 
\begin{equation} 
\label{q3} 
q= 1+ \frac{\nu}{1+\alpha} . 
\end{equation} 
  
	Note that setting $\nu = 1$ we recover the result of the dynamic  
treatment of 
Eq. (\ref{q2}), while the choice $\nu = 2$ yields Eq. (\ref{q1}). As  
earlier pointed out, in the  
original work of \cite{[ZA95]}, the assumption 
was made that the energy of the diffusing particle 
fluctuates 
as a result of the interaction with a bath.  
This physical  
condition corresponds to that of Ref.\cite{[PA97]}, a work based on the  
joint use of Tsallis generalized thermodynamics \cite{[T88]}, and  
N\'ose-Hoover (NH) dynamics \cite{[N84],[H85],[H91],[EM90]}. In the case  
of ordinary thermodynamics, the NH dynamics generates, with the minimum  
number of auxiliary equations, the canonical equilibrium distribution. The  
work of \cite{[PA97]} applies the NH dynamics to the case of non ordinary  
statistical mechanics, so as to make energy fluctuate as in the picture  
adopted by Alemany and Zanette \cite{[AZ94]}.  
 
	A final issue to discuss is now the thermodynamical 
significance of the generalized version of the Kolmogorov entropy,  
namely the trajectory entropy. 
We note that the function $\delta(t)$ of Eq. (\ref{deltax2})  
fits very well the corresponding prediction of Eq. (\ref{deltax}).  
We have also seen that 
the function $\psi(t)$ of Eq. (\ref{psi}) agrees, through Eq.  
(\ref{pipsi}), with the maximization of the Tsallis entropy, namely the 
distribution entropy. From the comparison of Eq. (\ref{psi}) with  
Eq. (\ref{deltax2}) we obtain 
\begin{equation} 
\label{keyrelation} 
\psi(t)= \kappa \delta(-\sigma(y_{0})t)  ~~~~~~~~~~(t> 0), 
\end{equation} 
with $\sigma(y_{0}) = {d^{z-1}}/{y_{0}^{z-1}}$ and $\kappa = ad^{z-1}$. 
A connection of this kind,  
between $\psi(t)$ and $\delta(t)$, with $\sigma$ independent of  
$y_{0}$, is expected on an intuitive ground.  
In fact, the lifetime of a laminar region  
is determined by the time making the distance between two  
trajectories, initially very close to one another, infinitely large 
and so exceeding the size of the laminar region. Note however 
that $\psi(t)$ is independent of $y_{0}$ and that $\delta(t)$ depends  
on it. This is not surprising, since $K_{q}$, as the conventional  
Kolmogorov entropy $K_{1}$, is a trajectory property. To 
make $K_{q}$ compatible with thermodynamics, a process forcing the 
single trajectory lose its dependence  
on its initial condition should be considered. The adoption 
of the distribution entropy seems to be more appropriate, to establish a  
connection between dynamics and mechanics, as a  
consequence of the fact that it explicitly involves a statistical  
average over many distinct trajectories, in accordance with the fact  
that also the dynamic derivation of the function $\psi(t)$  
is obtained\cite{[GT84]} by making averages over the trajectories 
injected with uniform probability into  
a given laminar region. This explains 
why the function $\psi(t)$ can be successfully determined using the 
entropic argument, namely from Eq.(\ref{pi2}) through Eq. (\ref{pipsi}),  
with no dependence on the initial conditions.

We conclude noticing that the microcanonical treatment of this letter 
implies \cite{[note3]} the mechanical energy to be kept fixed  
\cite{[AT98]}. The condition with the mechanical energy  
fluctuating, even if this is realized according to non-standard statistical  
mechanical prescriptions \cite{[ZA95]}, is equivalent to introducing  
probabilistic arguments, into a picture  
that, in principle, should remain mechanical at any step. In the  
present treatment the  
adoption of a stochastic assumption is confined to the chaotic  
transition from one laminar region to the other, and it is made  
apparently unnecessary by the adoption of distributions rather than  
trajectories.  
 
MB and AM thank the CNR for a  
fellowship and the dynamical  
systems group of the CNR Biophysics Institute of Pisa, Italy, for  
financial support, respectively.


\begin{thebibliography}{99}
\bibitem{[note1]} It has to be noticed that the L\'evy processes have been  
originally introduced as the most general solution of the  
Chapman-Kolmogorov equation \cite{[MS84]}. This means a completely  
probabilistic origin, with no indication on their dynamic origin. The  
discussion of their dynamic origin is much more recent, see, for instance,  
Refs. \cite{[TFWG94],[AGW96],[KZ93],[KZ94]}.
\bibitem{[note3]} The papers of Refs.\cite{[J95],[BMWG95]} are an  
illuminating example, although confined to the case of ordinary  
statistical mechanics, of a  
genuine derivation of thermodynamics from mechanics. According to  
\cite{[BMWG95]}, the derivation of ordinary Brownian motion from dynamics  
is equivalent to the dynamical derivation of canonical equilibrium. The  
ordinary thermal baths of the conventional derivation of Brownian motion  
are replaced by dynamical systems, assumed to be in a microcanonical  
condition determined by strong-chaos properties. Jarzinsky \cite{[J95]}  
adopts a theoretical perspective close to that of Ref. \cite{[BMWG95]}.  
The dynamic approach to anomalous diffusion of Refs.  
\cite{[AGW96],[KZ93],[KZ94]} is compatible with this microcanonical  
condition, and the search for the dynamic derivation of the entropic index  
$q\neq 1$ illustrated in this letter is inspired by the microcanonical  
perspective of \cite{[J95]} and \cite{[BMWG95]} as well as by the recent  
theoretical developments of \cite{[LT98],[TPZ97],[CLPT97]}.
\end{thebibliography}
\end{document}